\documentclass[aps,pre,twocolumn,preprintnumbers,amsmath,amssymb,nofootinbib]{revtex4-2}
\usepackage{epsfig}
\usepackage{epstopdf}
\usepackage{lipsum}
\usepackage{multirow}
\usepackage{boldline}
\usepackage{xcolor}
\usepackage{amssymb}
\usepackage{stackrel}
\usepackage[english]{babel}
\usepackage{tikz}
\usepackage{physics}
\newcommand{\overbar}[1]{%
	\tikz[baseline=(X.base)]{
		\node[inner sep=0pt] (X) {$#1$};
		\draw[line width=0.12ex] 
		([yshift=2.9ex,xshift=0.2em]X.south west) -- 
		([yshift=2.9ex,xshift=-0.2em]X.south east);
	}%
}

\newcommand{\leftrarrows}{\mathrel{\raise.75ex\hbox{\oalign{%
				$\scriptstyle\leftarrow$\cr
				\vrule width0pt height.5ex$\hfil\scriptstyle\relbar$\cr}}}}
\newcommand{\lrightarrows}{\mathrel{\raise.75ex\hbox{\oalign{%
				$\scriptstyle\relbar$\hfil\cr
				$\scriptstyle\vrule width0pt height.5ex\smash\rightarrow$\cr}}}}
\newcommand{\Rrelbar}{\mathrel{\raise.75ex\hbox{\oalign{%
				$\scriptstyle\relbar$\cr
				\vrule width0pt height.5ex$\scriptstyle\relbar$}}}}

\makeatletter
\def\leftrightarrowsfill@{\arrowfill@\leftrarrows\Rrelbar\lrightarrows}
\newcommand{\xleftrightarrows}[2][]{\ext@arrow 3399\leftrightarrowsfill@{#1}{#2}}
\makeatother

\usepackage{amssymb,amsmath}
\usepackage{mathtools} 


\usepackage{tikz}

\begin{document}
	
	\title{Universality classes in the time evolution of epidemic outbreaks on complex networks}
	\author{Mateusz J. Samsel, Agata Fronczak, Piotr Fronczak}
	\affiliation{Faculty of Physics, Warsaw University of Technology, Koszykowa 75, PL-00-662 Warsaw, Poland}
	\date{\today}
	
	\begin{abstract}
We investigate the full temporal evolution of epidemic outbreaks in complex networks, focusing on the susceptible-infected (SI) model of disease transmission. Combining theoretical analysis with large-scale numerical simulations, we uncover two universal patterns of epidemic growth, determined by the structure of the underlying network. In small-world networks, the prevalence follows a Gompertz-like curve, while in fractal networks it evolves according to Avrami-type dynamics—typical of spatially constrained systems. These regimes define distinct universality classes that remain robust across arbitrary transmission rates. Notably, our approach provides explicit analytical formulas for the global epidemic prevalence and class-specific scaling relations capturing its dependence on the transmission rate. We show that the commonly assumed early exponential growth occurs only in small-world networks, where it corresponds to the short-time approximation of the Gompertz function. In contrast, this exponential phase is entirely absent in fractal networks, where spreading is markedly slower and governed by different mechanisms. Our approach clarifies the structural origins of these contrasting behaviors and offer a unified framework for understanding epidemic dynamics across diverse network topologies.
\end{abstract}

\maketitle

\section{Introduction}
The spread of epidemics in complex networks~\cite{2022BookDorogovtsev} is a significant area of research with both theoretical and practical implications \cite{2015RevModPhys_PastorSatorras}. Theoretically, understanding the mechanisms governing the spread of infectious diseases sheds light on non-equilibrium dynamics and phase transitions.  Practically, reliable and accurate epidemic models are crucial for designing effective containment strategies, optimizing vaccination campaigns, and mitigating the impact of outbreaks in real-world populations.

The last two decades have brought significant progress   in understanding epidemic spreading on networks. Classical compartmental models, such as susceptible-infected-susceptible (SIS) and susceptible-infected-removed (SIR), have been extended to account for the heterogeneity of real-world contact patterns. Network-based approaches have revealed the importance of structural factors such as degree distributions, node-degree correlations, and even temporal patterns in shaping epidemic dynamics (see,~e.g.,~\cite{2001PRLPasotrSatorras, 2001PREPasotrSatorras, 2001PREMay, 2002PREBoguna, 2010PRLCastellano, 2013PRLBoguna, 2015PRXValdano}). In particular, the discovery that scale-free networks exhibit vanishing epidemic thresholds has had profound implications for public health policy and epidemic response strategies.

Despite these advances, several open questions remain. In particular, while extensive research has been devoted to steady-state properties and epidemic thresholds, much less attention has been given to the theoretical studies on the temporal evolution of outbreaks. The limited amount of research in this area (with only several exceptions, including~\cite{2002EPJBMoreno, 2004PRLBarthelemy, 2005JTeorBiolBarthelemy, 2007PhysAAvramov, 2010PREKarrer, 2014PRELokhov, 2020PRLMoore, 2022PREMerbis, 2024arxivCure, 2024PRLMoore}) is particularly concerning, especially since the speed of epidemic spreading is a fundamental aspect of disease dynamics. This factor determines how quickly an infection propagates through a network and how effectively interventions can be implemented.

In this paper, we address these issues by examining the full temporal evolution of epidemic outbreaks in complex networks using the most fundamental and relevant spreading process: the susceptible-infected (SI) model of disease transmission. We demonstrate that epidemic prevalence follows one of two distinct growth scenarios (i.e. universality classes), depending on the underlying network structure.

In the first scenario, which is typical for small-world networks \cite{1998NatureWatts,2003PRLCohen, 2004PREFronczak} such as classic random graphs and scale-free networks, the epidemic prevalence (fraction of infected nodes) follows a Gompertz growth curve \cite{1932PNASWinsor}: $\rho(t)\!\simeq\!1\!-\!const\,\exp(-\rho_0 m(t))$, where $\rho_0$ is the initial prevalence, and $m(t)\sim (q(R_0{-}1){+}1)^t$. Here, $q$ denotes the transmission rate, and $R_0$ is the basic reproduction number, representing the average number of secondary infections caused by a single individual in a fully susceptible population. In the second scenario, which is typical of fractal complex networks \cite{2005NatureSong, 2024SciRepFronczak} and systems with a clear spatial structure like regular grids, the prevalence follows the Avrami equation \cite{2023reviewShirzad}: $\rho(t)\!\simeq\!1\!-\!\exp(-const\,\rho_0 (q\,t)^d)$, where $d$ is a characteristic exponent depending on the dimensionality of the system.

Our results substantially enrich the current understanding of epidemic spreading in complex networks by explaining how the temporal evolution of outbreaks depends on network structure. Earlier approaches, particularly those based on mean-field theory and its heterogeneous variants, showed that during the early stages of an epidemic, the number of infections grows exponentially over time i.e., $\rho(t) \sim \exp(t/t_0)$, where $1/t_0=q(R_0{-}1)$ (see Chapter 7 in \cite{2022BookDorogovtsev} for a concise overview). We confirm this exponential behavior but show that it applies only to networks exhibiting the small-world effect. In such networks, the exponential growth corresponds to the lowest-order approximation of a more general growth function---the Gompertz curve---which accurately describes the full temporal course of the epidemic.

By contrast, we find that epidemic outbreaks in fractal complex networks follow fundamentally different dynamics, with no trace of exponential growth. In these systems, the spread is considerably slower and governed by Avrami-type behavior, which is characteristic of spatially extended systems. This distinction is especially important because many real-world complex networks, including social, technological, informational, and biological systems \cite{2005NatureSong,Gallos_2012,2024SciRepFronczak,2021Wen,2025arxivLepek,2025Makulski}, exhibit fractal properties. The SI model, which we analyze in this paper, provides a natural description not only of infectious disease transmission but also of information propagation.

The SI model is considered the most fundamental theoretical framework for assessing the impact of network topology on epidemic dynamics \cite{2004PRLBarthelemy, 2020PRLMoore, 2024arxivCure} and diffusion of information \cite{2016Zi}. We consider a population of $N$ individuals, each of whom can be in one of two discrete states: susceptible (S) or infected (I).We start with completely healthy population, in which at time $t=0$ infected individuals (so-called patient zeros) randomly appear, meaning that each node becomes infected with probability $\rho_0$, serving as the source of the epidemic. For $t>0$, the infection spreads iteratively as infected individuals transmit the disease to their nearest susceptible neighbors, who then continue the process. Once infected, nodes remain in this state until the epidemic fully saturates the population, at which point the process terminates. We track the epidemic's progression by counting the number of infected individuals $I(t)$, whose normalized density defines the epidemic prevalence, $\rho(t)=I(t)/N$.

To uncover the universal mechanisms underlying epidemic spreading in complex networks, we begin by analyzing the case of maximal transmission rate ($q = 1$), where an infected node at time $t$ transmits the disease to all its susceptible neighbors at the next time step (Section~\ref{beta1}). This idealized setting offers particularly transparent conditions that facilitate intuition and help reveal the core mechanisms driving the spread. The insights gained here form a conceptual basis for the more general case of arbitrary transmission probability $q \neq 1$, which we consider in Section~\ref{beta}, where each susceptible neighbor becomes infected independently with probability $q$. Importantly, this generalization shows that the emergence of two distinct universality classes characterizing temporal evolution of epidemic outbreak is driven purely by the network structure and not by the stochastic properties of the process itself.

\section{Maximal rate of epidemic transmission}\label{beta1}

\subsection{General equations for the local and global epidemic prevalence}\label{beta1A}

We begin our theoretical considerations with an expression describing the probability that node $i$ in the network is infected at time $t$, which we refer to as the local prevalence, $\rho_i(t)$. This probability can be calculated using the complement rule for the union of independent events, which states that the probability that at least one of several independent events occurs is equal to one minus the probability that none of them occurs (see Appendix~A).

Assuming the maximal rate of transmission, $q=1$, i.e., that each infected node always transmits the infection to all its susceptible neighbors in the next time step, the probability that node $i$ is not infected at time $t$ is equal to the probability that none of the nodes located within distance $t$ from $i$ was a patient zero. Therefore, the local prevalence is given by (cf.~Eq.~(\ref{Apx1}) in Appendix A):
\begin{equation}\label{rhoi00}
	\rho_i(t)=1-(1-\rho_0)^{m_i(t)},
\end{equation}
where $m_i(t)$ is the number of nodes within distance $t$ from node $i$. That is, $\rho_i(t)$ gives the probability that node $i$ is infected at time $t$, either because it was a patient zero itself or because the infection has reached it by time $t$ through a chain of successful transmissions initiated by some patient zero. In particular, since at distance zero there is only the node $i$ itself, we have $m_i(0) = 1$, and thus the local prevalence at time $t = 0$ simply equals the initial density of patient zeros: $\rho_i(0) = \rho_0$.

Although Eq.~(\ref{rhoi00}) has a very simple and self-explanatory form, in the following, in order to increase the analytical clarity of our derivations, we use its approximate form (cf.~Eq.~(\ref{Apx2}) in Appendix A):
\begin{equation}\label{rhoi0}
	\rho_i(t)\simeq 1-\exp(-\rho_0\,m_i(t)).
\end{equation} 
This approximation leads to more interpretable and elegantly structured expressions, while its error vanishes as $\rho_0^{\,2}m_i(t)$. For convenience, we also refer to the neighborhood of a node defined above and having size $m_i(t)$ as its susceptibility area. 

Averaging Eq.~(\ref{rhoi0}) over all nodes in the network yields theoretical expressions that describe the global epidemic prevalence:
\begin{equation}\label{rho0} 
	\rho(t)=\langle \rho_i(t) \rangle. 
\end{equation}
The scaling relations obtained in this way are then verified through numerical simulations, thereby confirming the hypotheses presented in this paper regarding the distinct universality classes of spreading processes in complex networks.

It is worth noting that Eq.~(\ref{rhoi0}), as formulated for systems with a graph-based structure, has a well-known continuous Euclidean counterpart:  
\begin{equation}\label{avrami00} 
	\rho_i(t)=1-\exp(-\rho_0\,c\,t^d), 
\end{equation}  
which can be obtained by substituting $m_i(t)\!=\!ct^d$ into~(\ref{rhoi0}), where $c$ is a constant parameter and $d$, the Avrami exponent, corresponds to the spatial dimension of the system. Eq.~(\ref{avrami00}) is widely recognized in physics as the Johnson-Mehl-Avrami-Kolmogorov equation, or simply the Avrami equation. In particular, in materials science, it is used to describe the kinetics of phase transformations, such as crystallization (see \cite{1998FanfoniReview} for a concise overview, and Chap.~9 in~\cite{2020bookCantor} for a historical perspective on the subject). Importantly, Eq.~(\ref{avrami00}) in this form can also be used to describe spreading processes in regular lattices, such as linear chains and square grids, where the spatial dimension corresponds to $d=1$ and $d=2$, respectively.

\subsection{Complex networks with the small-world property} 

One of the defining characteristics of complex networks is their scale-free nature \cite{1999ScienceBarabasi}, characterized by a power-law degree distribution, $P(k)\sim k^{-\gamma}$, in which a few highly connected nodes, known as hubs, play a crucial role in maintaining the network's overall connectivity. Another key property of complex networks, particularly relevant to the study of spreading phenomena, is the small-world effect \cite{1998NatureWatts}. This effect refers to the fact that the shortest path between any two nodes in such a network is relatively short compared to the network size, leading to the widespread use of the term 'small worlds' to describe these systems.

In the context of the results presented in this study, the above popular-science explanation of the small-world effect requires refinement. In network science, the term 'small-world networks' refers specifically to networks, or more precisely their synthetic models, in which the average shortest path length grows at most logarithmically with the network size \cite{2003PRLCohen, 2004PREFronczak}. This formal definition of small-worldness excludes fractal complex networks \cite{2005NatureSong}, in which the average shortest path scales as a power of the number of nodes \cite{2010PRLRozenfeld, 2024SciRepFronczak}. Naturally, individual realizations of fractal networks, for a fixed network size, may still exhibit short path lengths, making them 'small worlds' in the popular, non-technical sense of the term. However, introducing this distinction is essential for properly analyzing and interpreting the scaling relations governing epidemic spreading in different network topologies, as discussed further below.

In particular, as shown in \cite{2001PRENewman,2005PREHolyst}, the logarithmic scaling of the average path length in small-world networks arises because the number of nodes in the neighborhood of any given node grows exponentially with distance. This property is characteristic of many fundamental network models, including random graphs with arbitrary degree distributions (also the classic Erd\"os-R\'enyi random graphs) and various evolving network models (e.g., the seminal BA model~\cite{1999PhyABarabasi}). 

In such models, in the limit of large network size, the number of nodes within the susceptibility area of a node $i$ after $t$ time steps can be approximated as:
\begin{eqnarray}\label{swmit}
	m_i(t)&=&m_i(t, k_i, R_0-1)\\\nonumber &=&1{+}k_i{+}\dots{+}k_i R_0^{t-1}=1+k_i\frac{R_0^{t}{-}1}{R_0{-}1},
\end{eqnarray}
where $k_i$ is the degree of node $i$, and $R_0$ is the already identified reproduction number, corresponding to the average degree of a nearest neighbor minus one. This expression remains valid only as long as the infection radius $t$ is small compared to the network diameter.

By substituting Eq.~(\ref{swmit}) into~(\ref{rhoi0}), one obtains a Gompertz-like growth curve that describes the local epidemic prevalence in the universality class of small-world networks:
\begin{equation}\label{rhoisw}
	\rho_i(t)=1-\exp\left(-\rho_0\,m_i(t,k_i,R_0-1)\right).
\end{equation}

\begin{figure*}[t]
	\includegraphics[width=0.9\textwidth]{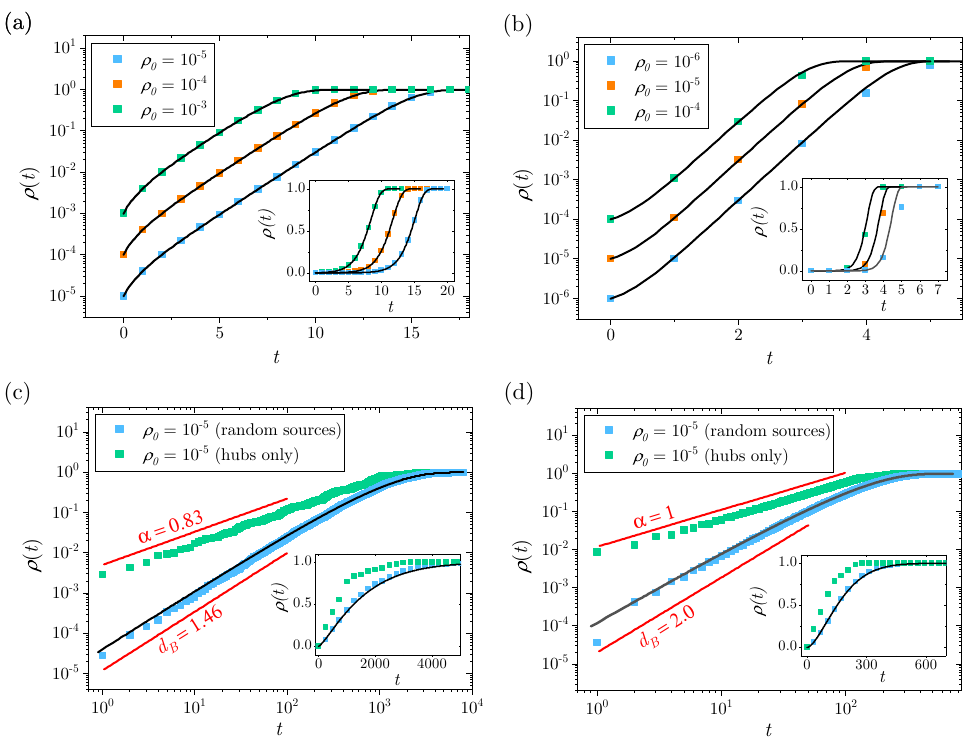}
	\caption{\textbf{Epidemic outbreak dynamics in the SI model with maximal transmission rate across various complex networks}: (a) $r$-regular random graph ($r=3$, $N=10^6$), (b) configuration model with a scale-free node degree distribution ($\gamma=3$, $k_0=5$, $N=10^6$)~\cite{2001PRENewman}, (c) Song-Havlin-Makse (SHM) model of fractal complex networks ($s=2$, $N\simeq 10^7$) \cite{2006NatPhysSong}, and (d) deterministic fractal network model known as ($u$,$v$)-flowers ($u=v=2$, $N\simeq 10^7$) \cite{2007NeeJphysRozenfeld}. Color-coded points represent numerical simulation results for different initial epidemic prevalences, averaged --$\,$in the case of 'random sources'$\,$-- over 50 different locations of the zero patients. Black solid lines indicate theoretical predictions based on the equations provided in the main text: Eqs.(\ref{rhoisw}) and(\ref{rhoSF}) for small-world networks in panels (a) and (b), respectively, and Eq.~(\ref{rhoF}) for fractal complex networks in panels (c) and (d). Each main panel presents data in a format adapted to sigmoidal growth curves in a given universality class. Red solid lines indicate theoretical predictions based on Eq.~(\ref{rhoif}) assuming patient zeros are network hubs. Red dashed lines indicate the slopes of the lines resulting from the approximation given by Eq.~(\ref{rhoFsim}). Insets display the same data on a linear scale. \label{figure}}
\end{figure*}

The simplest network model belonging to this universality class is the so-called $r$-regular random graph, in which all nodes have the same degree: $k_i=r$ and $R_0=r-1$. In the model, local infection prevalence (\ref{rhoisw}) is identical for all nodes and equal to the global prevalence (see Fig.~\ref{figure}(a)). 

In small-world networks where nodes have varying degrees, the time dependence of global epidemic prevalence can be obtained by averaging the local, degree-dependent prevalences (\ref{rhoisw}) over the node degree distribution. For example, in scale-free networks, with
\begin{equation}\label{PkSFsw}
P(k)\sim k^{-\gamma},
\end{equation} 
where $k\geq k_0$ and $R_0=\langle k^2\rangle/\langle k\rangle-1$ \cite{2001PRENewman}, this averaging yields (see Eq.~(\ref{AA1}) in Appendix~B for details):
\begin{eqnarray}\label{rhoSF}
	\rho(t)=1-(\gamma{-}1)e^{-\rho_0}E_{\gamma}\left(\rho_0(m_i(t,k_0,R_0{-}1){-}1)\right),
\end{eqnarray}
where $m_i(t,k_0,R_0)$ (\ref{swmit}) stands for the time-dependent size of the susceptibility area of the least connected nodes in the network and $E_\gamma(z)=\int_1^\infty e^{-zx}x^{-\gamma}dx$ is the exponential integral function (see Fig.~\ref{figure}(b)).


Referring to Eq.(\ref{rhoSF}), it is worth noting that when the argument of the exponential integral function is small compared to the order $\gamma$ of this function, then $E_\gamma(z)\simeq e^{-z}/(\gamma-1)$ \cite{1990expint} and Eq.~(\ref{rhoSF}) can be approximated by:
\begin{eqnarray}\label{rhoSFsim} 
	\rho(t)\simeq 1-\exp\left(-\rho_0\,m_i(t,k_0,R_0-1)\right). 
\end{eqnarray}
Remarkably, this approximation is often accurate not only in the early stage of an epidemic. This can be easily explained by noting that Eq.(\ref{rhoSFsim}) is equivalent to the expression describing the local prevalence of the least connected nodes, cf.~Eq.(\ref{rhoisw}), which are the most abundant in scale-free networks. The significance of this result lies, on the one hand, in highlighting the Gompertz-like time evolution of epidemic spread in scale-free networks, and on the other hand, in emphasizing that this behavior differs from a purely exponential growth, although such a growth of $\rho(t)$ at early times emerges naturally from Eq.~(\ref{rhoSFsim}) when the exponential function is approximated to leading order, i.e. $\rho(t)\simeq \rho_0\,m_i(t,k_0,R_0-1)$.

\subsection{Fractal complex networks} 

As already noted, fractal complex networks, although characterized by scale-free degree distributions (\ref{PkSFsw}), do not exhibit the small-world property. Another feature that distinguishes these networks from small-world complex networks is that, when covered with non-overlapping boxes, with the maximum distance between any two nodes in each box less than $l_B$, they exhibit power-law scaling~\cite{2005NatureSong}: 
\begin{equation}\label{Nblb}
	N_B(l_B)/N\sim l_B^{-d_B},
\end{equation}
where $N_B(l_B)$ is the number of boxes of a given diameter, and $d_B$ is the fractal (or box) dimension of the network. 

In \cite{2024SciRepFronczak}, it was shown that for fixed $l_B$, the box mass distribution follows a power law:
\begin{equation}\label{Pm}
	P(m)\sim m^{-\delta},
\end{equation}
for $m\geq m_0$, where $m_0\sim\langle m\rangle\sim l_B^{d_B}$ (\ref{Nblb}). This is due the scale-invariant properties of these boxes, whose masses depend not only on their diameter $l_B$, but also on the degrees $h_{i}$ of the best-connected nodes (local hubs) inside these boxes: 
\begin{equation}\label{mlb}
	m_i(l_B,h_i)\sim l_{B}^{\;\alpha}\,h_i^{\;\beta}\,,
\end{equation} 
where $\alpha$ and $\beta$ are the so-called microscopic scaling exponents characterizing the local structure of fractal complex networks. The microscopic exponents $\alpha$ and $\beta$ (\ref{mlb}), which describe the local structure of fractal complex networks, and the macroscopic scaling exponents $d_B$ (\ref{Nblb}), $\gamma$ (\ref{PkSFsw}), and $\delta$ (\ref{Pm}), which characterize their global properties, are related to each other by the following scaling relations:
\begin{equation}\label{ab0}
	\alpha = \frac{\delta - 2}{\delta - 1} d_B, \quad \text{and} \quad \beta = \frac{\gamma - 1}{\delta - 1}.
\end{equation}

The scale-invariant structure of fractal complex networks results in a fundamentally different kinetics of epidemic spreading compared to small-world networks \cite{2004PRLBarthelemy, 2005JTeorBiolBarthelemy}. Initially, the epidemic propagates within small boxes containing the patient zeros. As the infection saturates these boxes, they become macroscopic hotspots, driving the spread within progressively larger self-similar boxes to which they belong. 

In general, such boxes can be treated as susceptibility areas of their nodes, with the box mass (\ref{mlb}) corresponding to the size of this area, provided that $l_B\sim t$, i.e.
\begin{equation}\label{mitfrac}
	m_i(t)=m_i(t,h_i)\sim h_i^\beta t^\alpha.
\end{equation}
For this reason, given by Eq.~(\ref{rhoi0}), the local infection prevalence $\rho_i(t)$ in fractal complex networks does not depend on the degree $k_i$ of the considered node $i$ and on the basic reproduction number $R_0$, as is the case in small-world networks (\ref{rhoisw}), but rather on the degree $h_i$ of the local hub in the box to which it belongs, as well as on the scaling exponents $\alpha$ and $\beta$ that characterize the local structure of fractal complex networks:
\begin{equation}\label{rhoif}
	\rho_i(t)=1-\exp\left(-\rho_0\,m_i(t,h_i)\right).
\end{equation} 
It is worth noting that the resulting expression shows the Avrami-like time dependence, cf.~(\ref{avrami00}) and~(\ref{rhoif}), in contrast to the Gompertz-like growth (\ref{rhoisw}), which is typical of small-world networks. 

To obtain the global epidemic prevalence in fractal complex networks, the local prevalence $\rho_i(t)$ must be averaged over all nodes in the network. However, since in Eq.(\ref{rhoif}) the degree $h_i$ of the local hub may change over time---a point we address later---it is more convenient to begin with the general equation for $\rho_i(t)$, Eq.~(\ref{rhoi0}), and average it over the probability distribution of nodes assigned to boxes of a given mass, cf.~Eq.~(\ref{Pm}):
\begin{equation}\label{Pim}
	P_i(m)=\frac{m}{\langle m\rangle}P(m)\sim m^{1-\delta},
\end{equation}
where $m\geq m_0(t)$ with 
\begin{equation}\label{fracm0it}
	m_0(t)\sim t^{d_B}
\end{equation}
standing for the smallest box mass of diameter $l_B\sim t$,  which, through a linear dependence on the average box size $\langle m\rangle$, cf.~Eq.~(\ref{Nblb}), exhibits a power-law dependence on time with an exponent equal to the box dimension of the network. 

Following this approach yields the expression (see Eq.~(\ref{AA2}) in Appendix~B for details):
\begin{eqnarray}\label{rhoF}
	\rho(t)=1-(\delta-2)E_{\delta-1}\left(\rho_0\,m_0(t)\right),
\end{eqnarray}
which closely resembles Eq.~(\ref{rhoSF}) in form, yet fundamentally differs due to the power-law dependence of $m_0(t)$ on time, as opposed to the exponential dependence of $m_i(t,k_0,R_0)$ in the former (see Fig.~\ref{figure}(c,d)). In addition, using the same reasoning that led to Eq.~(\ref{rhoSFsim}), the above expression (\ref{rhoF}) can be approximated by the Avrami-equivalent growth function: 
\begin{eqnarray}\label{rhoFsim}
	\rho(t)\simeq 1-\exp\left(-\rho_0\,m_0(t)\right),
\end{eqnarray}
which, for the same reasons as Eq.~(\ref{rhoSFsim}), often holds well across the entire range of temporal variability.

Finally, an important remark should be made about the expression~(\ref{rhoif}) and the aforementioned time dependence of degrees $h_i$ of local hubs. This effect occurs when boxes with small diameters and low-degree local hubs become, over time, part of larger-diameter boxes that often contain higher-degree local hubs. As a result, the Avrami exponent, which characterizes the time dependence of local prevalence, is not simply equal to the scaling exponent $\alpha$, but is usually higher.

In particular, the effect described above makes the kinetics of epidemic spreading dependent on the strategy of selecting zero patients. For example, when they are selected from among the global hubs --$\,$which are local hubs in the boxes to which they belong, regardless of box diameters$\,$-- then the Avrami exponent characterizing the global epidemic prevalence is indeed equal to $\alpha$ (see Fig.~\ref{figure}(c,d)). This result can easily be deduced from Eq.~(\ref{rhoif}), which no longer describes the local prevalence, but --$\,$due to the fact that all boxes have similar hubs$\,$--  the global one. On the other hand, when patient zeros are selected randomly from among all network nodes, as described by Eq.~(\ref{rhoF}), its value is greater then $\alpha$ and closer to $d_B$ (\ref{rhoFsim}).

\begin{figure*}[t]
	\includegraphics[width=0.9\textwidth]{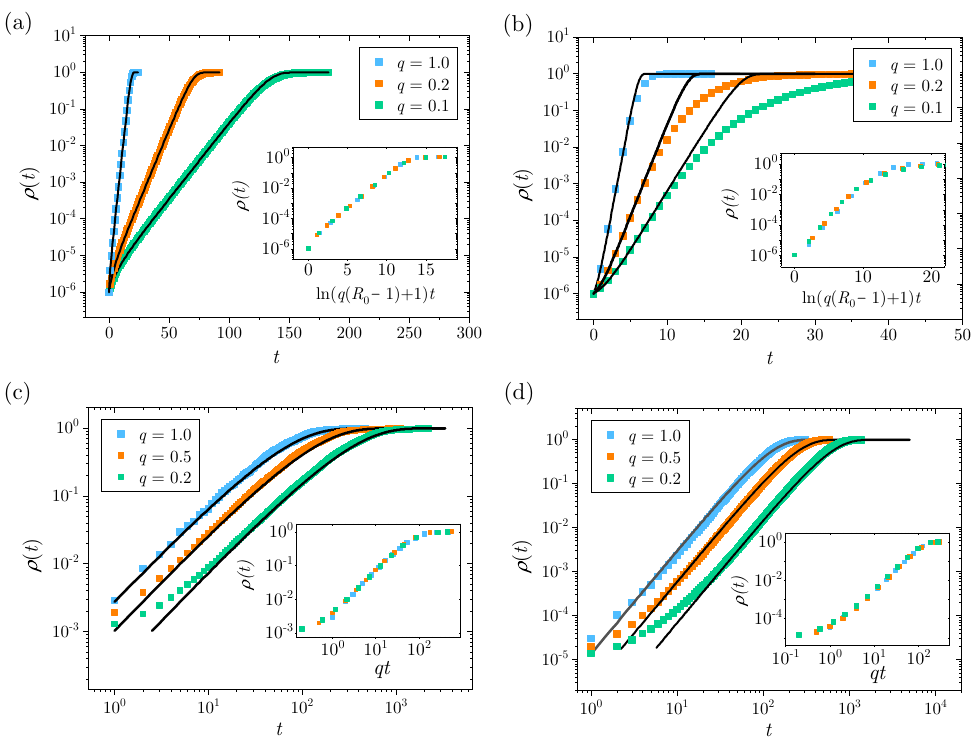}
	\caption{\textbf{Epidemic outbreak dynamics in the SI model with arbitrary transmission rate $q$ across various complex networks:} (a) r-regular random graph ($r = 3$, $N = 10^6$, $\rho_0=10^{-6}$), (b) configuration model with a scale-free node degree distribution ($\gamma = 3$, $k_0 = 2$, $N=10^6$, $\rho_0=10^{-6}$), (c) Song-Havlin-Makse (SHM) model of fractal complex networks ($s = 2$, $N = 15626$, $\rho_0=10^{-3}$), and (d) deterministic fractal network model known as $(u,v)$-flowers ($u = 4$, $v = 2$, $N = 223950$, $\rho_0=10^{-5}$). Color-coded points represent numerical simulation results for different transmission rates, averaged over 100 runs. Black solid lines indicate theoretical predictions based on Eq.~(\ref{rhoi0q}). Insets display the same data on a rescaled time scale according to Eq.~(\ref{swmitq}) (top row) and Eq.~(\ref{fracm0tq}) (bottom row).\label{fig2}}
\end{figure*}

\section{Arbitrary rate of epidemic transmission}\label{beta}

\subsection{General formulation for arbitrary transmission rate}\label{betaA}

In the previous section, we characterized the dynamics of epidemic spreading under the assumption of maximal transmission rate, $q = 1$, where an infected node transmits the disease to all of its susceptible neighbors in the next time step. Here, we extend the analysis to arbitrary transmission rates $q \neq 1$, where $q$ represents the probability that an infected node transmits the disease to a susceptible neighbor in a single time step. Despite this generalization, the qualitative nature of the spreading dynamics remains unchanged: small-world networks still follow Gompertz-like growth, while fractal networks exhibit Avrami-type behavior.

This robustness, understood as the structural invariance of the analytical expressions for epidemic prevalence stems from the fact that the generalized expression for local prevalence, derived in detail in Section~\ref{betaB} below,~i.e.
\begin{equation}\label{rhoi0q}
	\rho_i(t|q) = 1 - \exp(-\rho_0\,m_i(t|q)),
\end{equation}
preserves the same form as the Eq.~(\ref{rhoi0}) that underpinned all the earlier results.

However, the key difference between the above equation (\ref{rhoi0q}) and Eq.~(\ref{rhoi0}) lies in the interpretation of the argument of the exponential function. While $m_i(t)$ in Eq.~(\ref{rhoi0}) simply counts the number of nodes within a distance $t$ from node $i$, its generalized counterpart $m_i(t|q)$ in Eq.~(\ref{rhoi0q}) accounts for the probabilistic nature of transmission and quantifies the effective susceptibility area---that is, the expected number of nodes that could have infected node $i$ by time $t$. Although the topological neighborhood of node $i$ remains the same, the transmission probability $q$ modifies how each layer in that neighborhood contributes to the overall infection risk. As a result, $m_i(t|q)$ can be seen as a rescaled version of the original susceptibility area, reflecting not just who is connected, but how likely transmission is to occur across those connections. 

As discussed above, the structural invariance of the equations for local and global epidemic prevalence across different universality classes implies that the function $m_i(t|q)$ must follow certain scaling laws characteristic of each class. In Section~\ref{betaC} below, we show that this is indeed the case. For small-world networks, the effective susceptibility area scales as (cf.~Eq.~(\ref{swmit})):
\begin{equation}\label{swmitq}
	m_i(t,k_i,R_0-1|q)=m_i(t, q\,k_i, q\,(R_0-1)),
\end{equation}
while for fractal networks, it obeys the following scaling laws (cf.~Eqs.~(\ref{mitfrac}) and~(\ref{fracm0it})):
\begin{eqnarray}\label{fracmitq}
	m_i(t,h_i|q) &\simeq& m_i(q\,t, h_i),\\\label{fracm0tq}
	m_0(t|q) &\simeq& m_0(q\,t).
\end{eqnarray}

Substituting the Eqs.(\ref{swmitq})--(\ref{fracm0tq}), into the prevalence equations Eqs.~(\ref{rhoisw})--(\ref{rhoSFsim}) from Section II for small-world networks and Eqs.~(\ref{rhoif})--(\ref{rhoFsim})  for fractal networks we obtain a complete and consistent description of epidemic dynamics for any transmission probability. As demonstrated in Fig.~\ref{fig2}, the theoretical predictions remain in very good agreement with numerical simulations.

\subsection{Probabilistic justification of the generalized equation for local prevalence}\label{betaB}

To justify Eq.~(\ref{rhoi0q}) for the local epidemic prevalence $\rho_i(t|q)$ in the SI model with arbitrary transmission rate $q$, we apply the complement rule for the union of independent events (see Appendix A), as we did in Section~\ref{beta1A} to derive Eq.~(\ref{rhoi0}) for the special case $q = 1$.

Recall that $\rho_i(t|q)$ denotes the probability that node $i$ is infected at time $t$. This infection may have occurred either at time $\tau = 0$, meaning node $i$ was one of the patient zeros or at some later time $\tau = 1,2,\dots,t$ as a result of a transmission chain originating from a patient zero located at a distance $x \leq \tau$ (and no greater than $t$) from node $i$.

While the exact complement rule requires computing one minus the product of the probabilities that none of the independent infection events occur, we instead adopt the Poisson approximation (see Eq.~(\ref{Apx2}) in Appendix A). This approximation is appropriate when individual infection probabilities are small and allows us to estimate $\rho_i(t|q)$ as the sum of the probabilities of each infection path, treated separately. Importantly, this sum corresponds directly to the argument of the exponential function in Eq.~(\ref{rhoi0q}); up to the constant factor $\rho_0$, it defines the effective susceptibility area $m_i(t|q)$.

To calculate this sum, we begin by analyzing the infection of node $i$ occurring at time $\tau = 1,2,\dots,t$, initiated by a patient zero located at a distance $x = \tau, \tau+1,\dots,t$ from node $i$. Assuming that the infection spreads along the shortest path, the probability of such an event is given by:
\begin{equation}\label{help0}
	\rho_0\,q\,P_q(x{-}1|\tau{-}1) = \rho_0\,\binom{\tau{-}1}{x{-}1}q^{x}(1{-}q)^{\tau{-}x},
\end{equation}
where $\rho_0$ is the probability that the node at the beginning of the path is a patient zero, and $P_q(x{-}1|\tau{-}1)$ is the binomial probability of exactly $x{-}1$ successful transmissions (each occurring with probability $q$) within $\tau{-}1$ time steps. Importantly, Eq.~(\ref{help0}) captures the fact that the infection must travel successfully across $x{-}1$ links before reaching node $i$, with the final successful transmission (i.e., $x$-th) occurring precisely at time $\tau$. The remaining $\tau{-}x$ time steps correspond to unsuccessful edge-based transmission attempts, each occurring with probability $1{-}q$, and interspersed among the successful ones.

Finally, by summing the probability $\rho_0$ that node $i$ is a patient zero together with the probabilities of all individual infection paths described by Eq.~(\ref{help0}), we recover Eq.~(\ref{rhoi0q}), where the argument of the exponential function defines the effective susceptibility area:
\begin{equation}\label{help2}
	m_i(t|q) = 1 + q\sum_{\tau=1}^t \sum_{x = 1}^{\tau} n_i(x)\,P_q(x{-}1|\tau{-}1),
\end{equation}
where $n_i(x)$ denotes the number of nodes at distance $x$ from node $i$, i.e. such that: $1+\sum_{x=1}^tn_i(x)=m_i(t)$.

\subsection{Scaling relations for the effective susceptibility area}\label{betaC}

The three scaling relations for effective susceptibility areas—Eqs.~(\ref{swmitq})--(\ref{fracm0tq})—are special cases of the general formulation given in Eq.~(\ref{help2}), each resulting from a specific assumption about the network structure, reflected in $n_i(x)$ within the susceptibility area.

In particular, Eq.~(\ref{help2}) reduces to the original expression $m_i(t) = m_i(t|1)$ when $q = 1$, as can be seen by noting that $P_1(x{-}1|\tau{-}1) = \delta_{x,\tau}$, where $\delta_{x,\tau}$ denotes the Kronecker delta. This implies that, for $q = 1$, only patient zeros located exactly at distance $\tau$ from node $i$ can infect it at time $\tau$, thereby reducing the generalized formulation introduced in Section~\ref{betaA} to the original formulation used in Section~\ref{beta1A}.

Correspondingly, Eq.~(\ref{swmitq}) is obtained by substituting into Eq.~(\ref{help2}) the characteristic structure of small-world networks, where the number of nodes at distance $x$ from a given node grows exponentially as $n_i(x) = k_i R_0^{x - 1}$ (\ref{swmit}) (see Eq.~(\ref{AB1}) in Appendix~C for details). 

Finally, the fractal case, as described by Eqs.~(\ref{fracmitq}) and~(\ref{fracm0tq}), requires a more nuanced analysis.  In particular, to justify the first of these equations it is assumed that $n_i(x)\simeq \dv*{m_i(x{,}h_i)}{x}=c\,\alpha\, h_i^\beta x^{\alpha-1}$ (\ref{mitfrac}), where $c=const$. Then valuating the resulting sum in Eq.~(\ref{help2}) involves a mean-field approximation applied to the binomial distribution over distances. Specifically, to get Eq.~(\ref{fracmitq}) one approximates $\langle x^\alpha \rangle$ by $\langle x \rangle^\alpha$ (see Eq.~(\ref{AB2}) in Appendix~C for details). 

\section{Summary and concluding remarks} 
In this study, we investigate the temporal evolution of epidemic outbreaks on complex networks, especially those with scale-free degree distributions. We identify two distinct universality classes that govern the dynamics of those outbreaks. In small-world networks, including the configuration model and the seminal Barabási–Albert model, the prevalence of infection follows a Gompertz-like growth curve. In contrast, in fractal complex networks with a well-defined box-counting dimension, the prevalence evolves according to an Avrami-like time dependence, typical of spatially constrained systems.
	
These insights refine and broaden existing knowledge about epidemic spreading in complex networks. While most prior studies have focused on small-world structures and described early-time dynamics in terms of exponential growth derived from mean-field approximations, we show that this growth corresponds to a short-time limit of the Gompertz function, and we provide a full analytical description that captures the entire course of the epidemic. Crucially, we demonstrate that this exponential phase is entirely absent in fractal networks, where spreading is significantly slower and governed by fundamentally different underlying mechanisms. 
	
Beyond advancing existing theory, our results offer a new perspective on a key issue: the exceptional vulnerability of scale-free networks to epidemics, which is typically attributed to the absence of epidemic thresholds in such systems \cite{2001PRLPasotrSatorras, 2001PREPasotrSatorras}. This widely accepted view stands in contrast to a handful of noteworthy reports in the literature suggesting the existence of non-zero thresholds in fractal scale-free networks \cite{2008JStatMechZhang, 2021FractalsNian}. Our findings reconcile these observations by showing that epidemic dynamics in scale-free networks with small-world properties differ fundamentally from those in their fractal counterparts. The scale-free nature of a network, when defined solely by its degree distribution, should therefore not be treated as the sole predictor of epidemic behavior, especially when considering such critical issues as prevention strategies, early-warning indicators, resilience assessment, or resource allocation.
	
This distinction between small-world and fractal complex networks becomes even more significant in light of empirical studies revealing that fractal structures are not rare exceptions. Although small-world networks dominate the mainstream discourse, many real-world systems—including the World Wide Web, the Internet, and various biological and social networks—exhibit fractal features \cite{2005NatureSong, Gallos_2012, 2024SciRepFronczak, 2021Wen, 2025arxivLepek, 2025Makulski}. For instance, hierarchical social networks may appear to belong to the small-world class due to long-range connections, yet they possess an underlying fractal skeleton. This observation also applies to many networks related to urban geography, such as road networks of major cities or the spatial distribution of urban populations, which have been shown to exhibit approximate fractal structure. These systems are of particular importance in modeling spreading phenomena, including epidemic outbreaks and information diffusion in metropolitan areas. These findings emphasize the importance of incorporating geometric constraints into models of spreading phenomena. Moreover, they suggest that such constraints may fundamentally affect the behavior of more realistic epidemic models (e.g., SIS or SIR), potentially altering the emergence of epidemic thresholds and the nature of transitions to endemic states. This highlights the need for further theoretical and numerical research on spreading processes in geometrically constrained complex networks, where traditional assumptions, such as the primacy of degree distribution, may no longer be sufficient.

A particularly noteworthy recent contribution in this direction is the study by Moore et al.~\cite{2024PRLMoore}, which explicitly incorporates network dimensionality into the modeling of epidemic dynamics. The authors propose a quasi-analytical iterative method in which the number of infected nodes at time \( t \) is reconstructed by assuming that the infection propagates within a ball-like region centered at the origin. The effective radius of this region is determined by the current number of infected nodes and the correlation dimension of the network. The number of new infections is then estimated by counting the susceptible nodes at distance \( r + 1 \) and updating the total accordingly. This approach offers an elegant and efficient procedure to predict epidemic trajectories based solely on structural characteristics and the current epidemic state.

Although the method proposed by Moore is undoubtedly promising, its applicability relies on the assumption that the studied network possesses a well-defined metric structure—most importantly, a meaningful correlation dimension. Our additional analyses (see Supplementary materials) suggest that, while the correlation dimension is well defined in fractal networks, its definition for small-world networks where numerous long-range connections disrupt the underlying geometry appears problematic. This explains why Moore’s method successfully reproduces the short and mid-term epidemic dynamics in fractal networks with results consistent with our scaling approach but fails to apply in small-world structures.

Despite the essential differences between these two approaches---one based on a quasi-analytical iterative scheme, the other on scaling theory---their agreement in the case of fractal networks highlights a shared message: the geometric structure of complex networks, especially their metric dimension, plays a key role in shaping the temporal evolution of spreading processes. Recognizing this fact and consistently incorporating it into theoretical models and numerical simulations is essential for developing more accurate and universal descriptions of real-world epidemic phenomena.

\section*{Acknowledgments} Research was funded by Warsaw University of Technology within the Excellence Initiative: Research University (IDUB) programme.

\section*{Appendix A}

Let $A_1, A_2, \dots, A_n$ be independent events, and define $A=\bigcup_{j=1}^n A_j$ as the event that at least one of them occurs. The complement rule for the union of independent events states that:
\begin{equation}\label{Apx1}
	P\left(A \right)=1-P\left(\bigcap_{j=1}^n \overbar{A_j} \right) = 1-\prod_{j=1}^n \left( 1 - P(A_j) \right),
\end{equation}
where $\overbar{A_j}$ denotes the event that $A_j$ does not occur. 

If all $P(A_j)$ are small (i.e., $P(A_j){\ll}1$), the product in Eq.~(\ref{Apx1}) can be approximately calculated as:
\begin{equation}
\prod_{j=1}^n (1 - P(A_j)) \simeq \exp\left( -\sum_{j=1}^n P(A_j) \right).
\end{equation}
Thus, Eq.~(\ref{Apx1}) becomes:
\begin{equation}\label{Apx2}
	P\left( \bigcup_{j=1}^n A_j \right) \simeq 1 - \exp\left( -\sum_{j=1}^n P(A_j) \right).
\end{equation}
This is known as the Poisson approximation and is widely used when modeling rare, independent events.

\section*{Appendix B}

Below we provide step-by-step derivations of Eqs.~(\ref{rhoSF}) and (\ref{rhoF}) from the main text, which were omitted for conciseness.

Thus, we arrive at Eq.~(\ref{rhoSF}) through the following steps:
\begin{eqnarray}\label{AA1}
	\rho(t)&=&\int_{k_0}^{k_m} \rho_i(t)P(k_i)dk_i\\\nonumber&=&\int_{k_0}^{k_m}\!\! \left(\!1\!-\!\exp\left(\!-\rho_0\left(\!1\!+\!\frac{R_0^{\,t}\!-\!1}{R_0\!-\!1}k_i\!\right)\!\right)\!\right) \frac{(\gamma\!-\!1)k_0^{\,\gamma\!-\!1}}{k_i^{\,\gamma}}dk_i\\\nonumber&=&1-(\gamma-1)e^{-\rho_0}\int_{1}^{\frac{k_m}{k_0}} \exp\left(-\rho_0 k_0\frac{R_0^{\,t}-1}{R_0-1} x \right)\frac{dx}{x^{\,\gamma}}\\\nonumber &\simeq &1-(\gamma-1)e^{-\rho_0}E_{\gamma}(\rho_0(\,m_i(t,k_0)-1)), 
\end{eqnarray}
where:
\begin{itemize}
	\item $k_0$ and $k_m$ represent the degree of the least and the most connected node in the network, respectively, 
	\item $P(k_i)=(\gamma-1)k_0^{\,\gamma-1} k_i^{\,-\gamma}$ is the node degree distribution in scale-free networks,
	\item $m_i(t,k_0)=1+k_0(R_0^{\,t}-1)/(R_0-1)$ stands for the time-dependent size of the susceptibility area of the least connected nodes, 
	\item $E_\gamma(z)=\int_1^\infty e^{-zx}x^{-\gamma}dx$ is the exponential integral function. 
\end{itemize}

Similarly, the derivation of Eq.~(\ref{rhoF}) proceeds as follows:
\begin{eqnarray}\label{AA2}
	\rho(t)&=&\int_{m_0}^{m_m} \rho_i(t)P_i(m)dm\\\nonumber &=&\int_{m_0}^{m_m}\left(1-\exp\left(-\rho_0\,m\right)\right) \frac{m}{\langle m\rangle} P(m)dm\\\nonumber &=&1-\frac{(\delta-1)m_0^{\,\delta-1}}{\langle m\rangle} \int_{m_0}^{m_m}\exp\left(-\rho_0\,m\right)\frac{dm}{m^{\delta-1}}\\\nonumber &=&1-(\delta-2)\int_{1}^{\frac{m_m}{m_0}}\exp\left(-\rho_0\,m_0\,x\right)\frac{dx}{x^{\delta-1}}\\\nonumber &\simeq& 1-(\delta-2)E_{\delta-1}(\rho_0\,m_0(t)),   
\end{eqnarray}
where:
\begin{itemize}
	\item $m_0$ and $m_m$ denote the smallest and the largest possible sizes of the susceptibility area, respectively, assuming that in fractal networks, these areas correspond to self-similar boxes, with their sizes representing the masses of these boxes,
	\item $P_i(m)\!=\!\frac{m}{\langle m\rangle}P(m)$ is the probability that the node $i$ belongs to the box of mass $m$, where $P(m)\!=\!(\delta\!-\!1)m_0^{\delta-1} m^{-\delta}$ is the box mass distribution,
	\item $\langle m\rangle=\int_{m_0}^{m_m}mP(m)dm=m_0\frac{\delta-1}{\delta-2}$ represents the average box mass, which, according to the power-law scaling of the number of boxes in fractal networks, varies over time, as: $\langle m\rangle=N/N_B(t)\sim t^{d_B}$ (see explanation in the main text),   
	\item $E_{\delta-1}(z)=\int_1^\infty e^{-zx}x^{-\delta-1}dx$ is, as before, the exponential integral function.
\end{itemize}

\section*{Appendix C}

The derivation of Eq.~(\ref{swmitq}) proceeds as follows:
\begin{align}\label{AB1}
	m_i(t,k_i,&R_0-1|q)=\\\nonumber
	&1+qk_i\sum_{\tau=1}^t \sum_{x=1}^{\tau}\binom{\tau{-}1}{x{-}1}(R_0q)^{x-1}(1{-}q)^{\tau{-}x}=\\\nonumber
	&1+qk_i\sum_{\tau=1}^t(q(R_0-1)+1)^{\tau{-}1}=\\\nonumber
	&1+qk_i\frac{(q(R_0-1)+1)^t-1}{q(R_0-1)}\stackrel{Eq.(\ref{swmit})}{=}\\\nonumber
	&m_i(t,q\,k_i,q\,(R_0-1)).
\end{align}

The derivation of Eq.~(\ref{fracmitq}) proceeds as follows:
\begin{align}\label{AB2}
	m_i(t,&h_i|q)=\\\nonumber
	&1+C\alpha h_i^\beta\sum_{\tau=1}^t\sum_{x=1}^{\tau}x^{\alpha{-}1}\binom{\tau{-}1}{x{-}1}q^{x}(1{-}q)^{\tau{-}x}=\\\nonumber &1+C\alpha h_i^\beta\sum_{\tau=1}^t\frac{1}{\tau}\sum_{x=1}^{\tau}x^\alpha\binom{\tau}{x}q^{x}(1{-}q)^{\tau{-}x}=\\\nonumber
	&1+C\alpha h_i^\beta\sum_{\tau=1}^t\frac{1}{\tau}\langle x^\alpha\rangle\simeq 1+C\alpha h_i^\beta\int_{1}^t\frac{1}{\tau}(\tau q)^\alpha d\tau\simeq\\\nonumber
	&1+Ch_i^\beta (qt)^\alpha\stackrel{Eq.(\ref{mitfrac})}{=}\\\nonumber
	&m_i(q\,t,h_i).
\end{align}

\end{document}